\documentclass{emulateapj}
\usepackage{natbib}
\usepackage{graphicx}
\usepackage{bm}

\usepackage[hyperfootnotes=false,naturalnames=true,letterpaper,pdfstartview=FitH,pdfpagemode=none]{hyperref}
\hypersetup{pdfauthor={Dan Coe}, pdftitle={Fisher Matrices}}

\slugcomment{Version 1}

\newcommand\p{\partial}

\newcommand\dxa{\displaystyle \frac{\p x}{\p a}}
\newcommand\dxb{\displaystyle \frac{\p x}{\p b}}
\newcommand\dxc{\displaystyle \frac{\p x}{\p c}}
\newcommand\dya{\displaystyle \frac{\p y}{\p a}}
\newcommand\dyb{\displaystyle \frac{\p y}{\p b}}
\newcommand\dyc{\displaystyle \frac{\p y}{\p c}}
\newcommand\dza{\displaystyle \frac{\p z}{\p a}}
\newcommand\dzb{\displaystyle \frac{\p z}{\p b}}
\newcommand\dzc{\displaystyle \frac{\p z}{\p c}}

\newcommand\pxx{\displaystyle \frac{\p^2}{\p x^2}}
\newcommand\pyy{\displaystyle \frac{\p^2}{\p y^2}}
\newcommand\pxy{\displaystyle \frac{\p^2}{\p x \p y}}

\newcommand\dchisqxx{\displaystyle \frac{\p^2 \chi^2}{\p x^2}}

\newcommand\dchisqxy{\displaystyle \frac{\p^2 \chi^2}{\p x \p y}}
\newcommand\dchisqx{\displaystyle \frac{\p \chi^2}{\p x}}

\newcommand\domOm{\displaystyle \frac{\p \om}{\p \Om}}
\newcommand\domOL{\displaystyle \frac{\p \om}{\p \OL}}
\newcommand\domh{\displaystyle  \frac{\p \om}{\p h}}
\newcommand\dOLOm{\displaystyle \frac{\p \OL}{\p \Om}}
\newcommand\dOLOL{\displaystyle \frac{\p \OL}{\p \OL}}
\newcommand\dOLh{\displaystyle  \frac{\p \OL}{\p h}}
\newcommand\dOkOm{\displaystyle \frac{\p \Ok}{\p \Om}}
\newcommand\dOkOL{\displaystyle \frac{\p \Ok}{\p \OL}}
\newcommand\dOkh{\displaystyle  \frac{\p \Ok}{\p h}}

\newcommand\Dxdx{\displaystyle \frac{\Dx}{\dx}}
\newcommand\Dydy{\displaystyle \frac{\Dy}{\dy}}

\newcommand\dx{\sigma_x}
\newcommand\dy{\sigma_y}
\newcommand\dxy{\sigma_{xy}}

\newcommand\Dx{\Delta x}
\newcommand\Dy{\Delta y}

\newcommand\Om{\Omega_m}
\newcommand\om{\omega_m}
\newcommand\OL{\Omega_\Lambda}
\newcommand\Ok{\Omega_k}

\begin{document}

\title{Fisher Matrices and Confidence Ellipses: 
A Quick-Start Guide and Software}

\author{Dan Coe}
\email{coe@caltech.edu}
\affil{
  Jet Propulsion Laboratory, California Institute of Technology, 
  4800 Oak Grove Dr, MS 169-327, Pasadena, CA 91109 
}

\begin{abstract}
Fisher matrices are used frequently in the analysis of 
combining cosmological constraints from various data sets.
They encode the Gaussian uncertainties of multiple variables.
They are simple to use, and I show how to get up and running with them quickly.
Python software is also provided.
I cover how to obtain confidence ellipses, add data sets, apply priors, marginalize, transform variables, and even calculate your own Fisher matrices.
This treatment is not new, 
but I aim to provide a clear and concise reference guide.
I also provide references and links to more sophisticated treatments and software.
\end{abstract}


\section{Outline}

I explain how to do/obtain the following with/from Fisher matrices:

\S\ \ref{S:confell}: Confidence Ellipses

\S\ \ref{S:manipulation}: Manipulation: \\
\indent \indent \indent \indent Marginalization, Priors, Adding Data Sets

\S\ \ref{S:calc}: How to Calculate your Own Fisher Matrices

\S\ \ref{S:trans}: How to transform variables

\S\ \ref{S:zp}: Dark energy pivot redshift

\S\ \ref{S:disc}: Discussion (brief) about what Fisher matrices are


\S\ \ref{S:software}: Software I've come across (including my own)

\S\ \ref{S:contribute}: How you can contribute to this paper





\section{Fisher Matrices $\Rightarrow$ Confidence Ellipses}
\label{S:confell}

\begin{figure*}
\includegraphics[width=0.45\hsize]{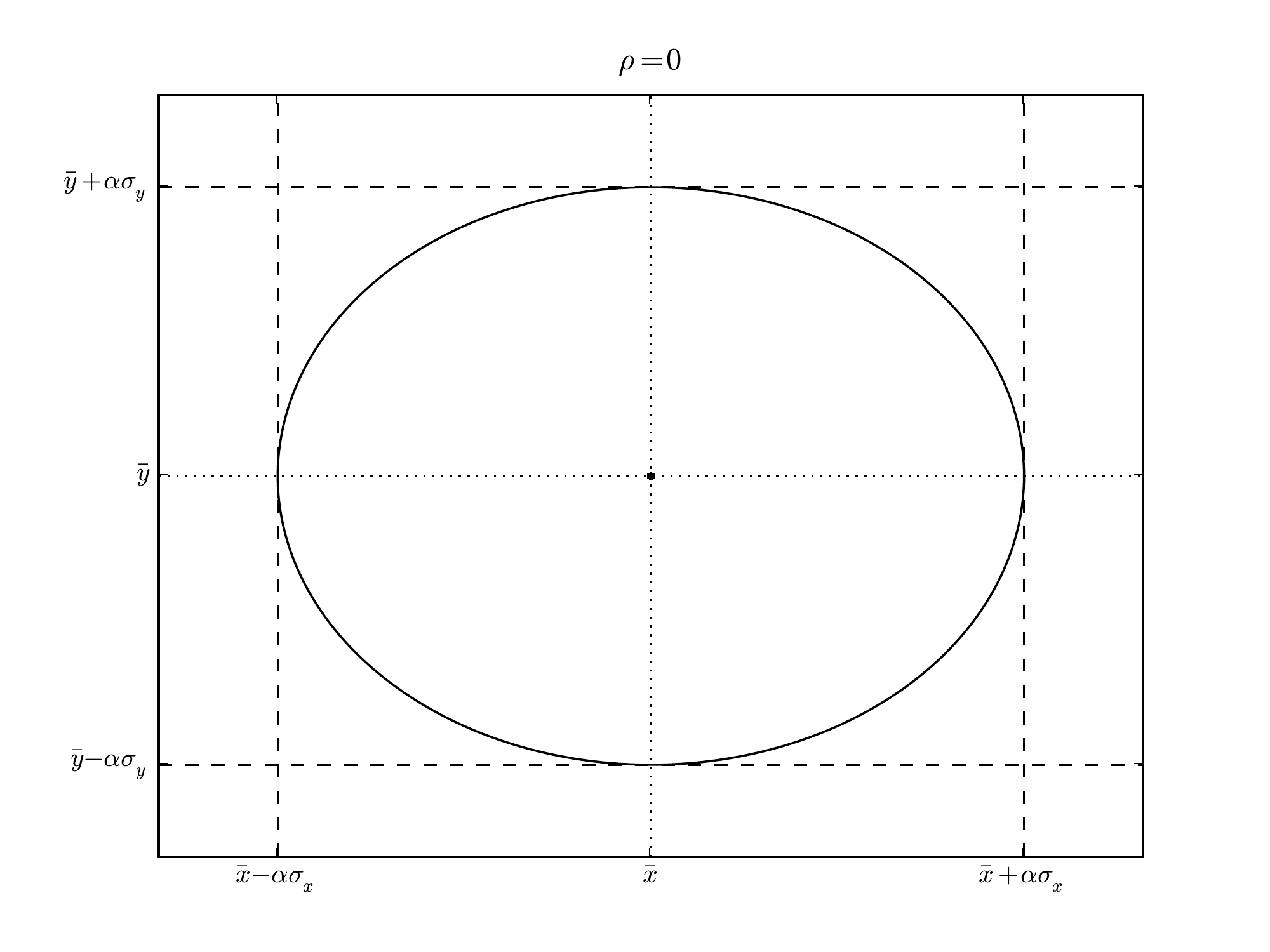}
\includegraphics[width=0.45\hsize]{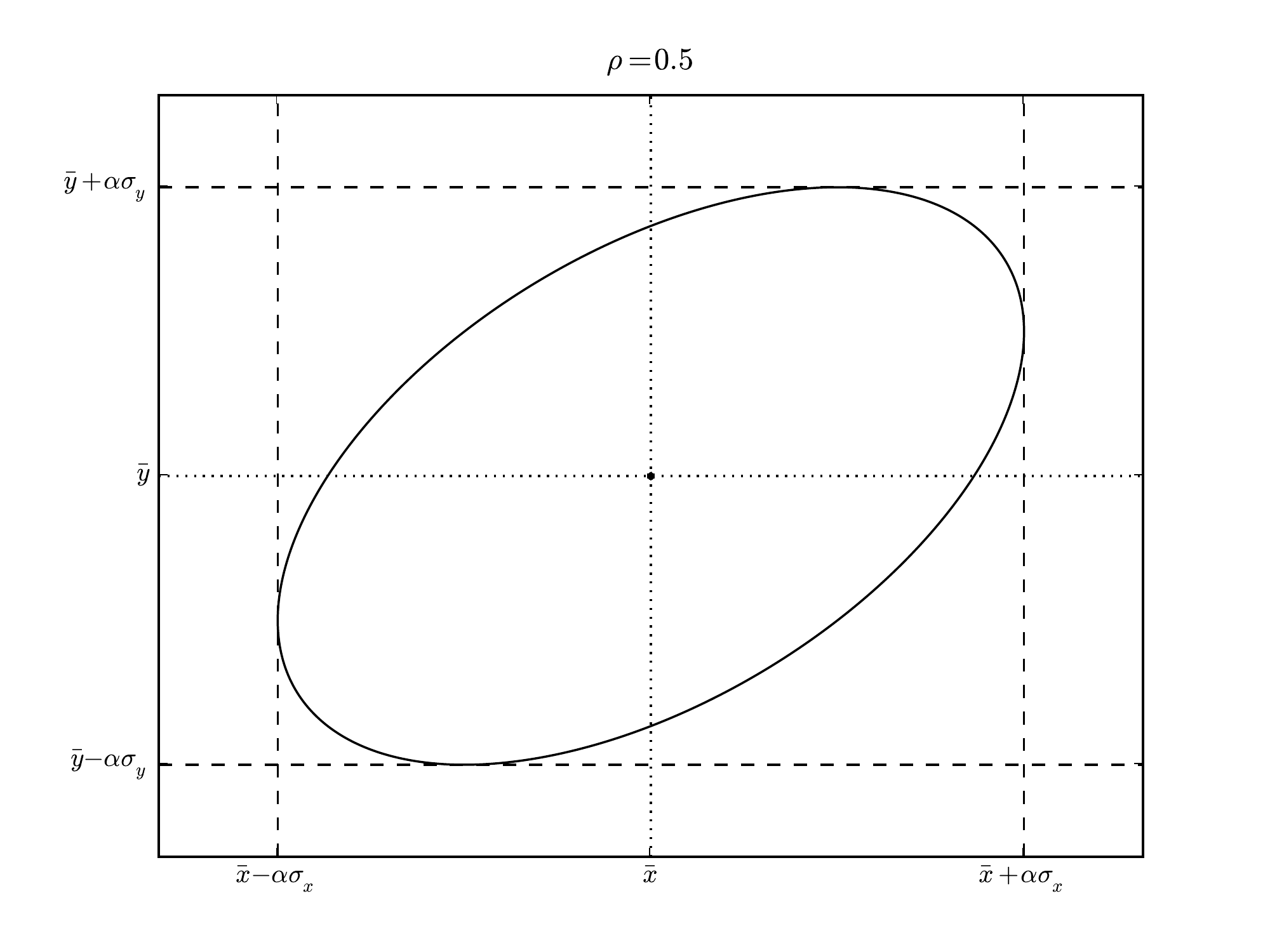}
\\
\includegraphics[width=0.45\hsize]{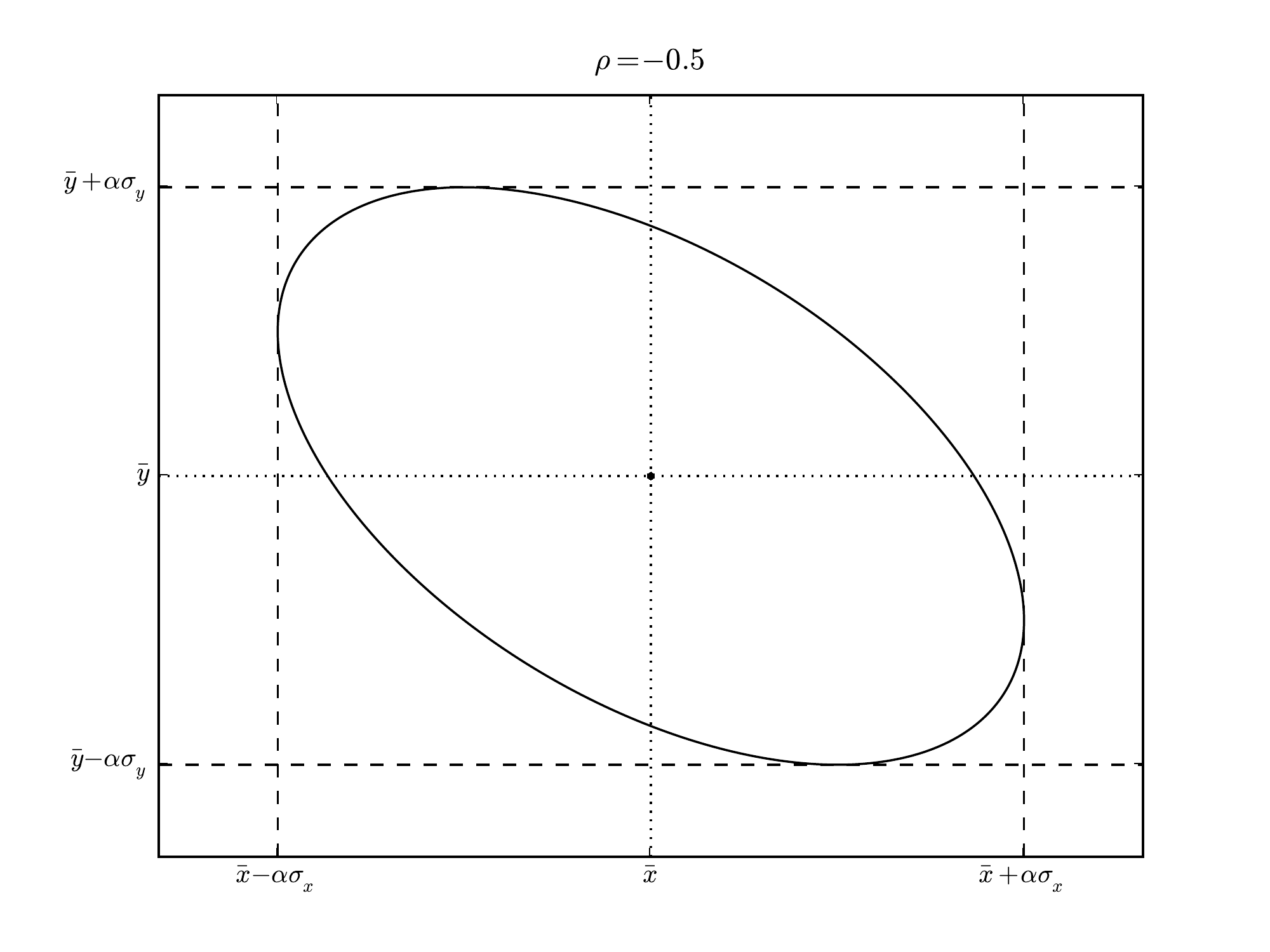}
\includegraphics[width=0.45\hsize]{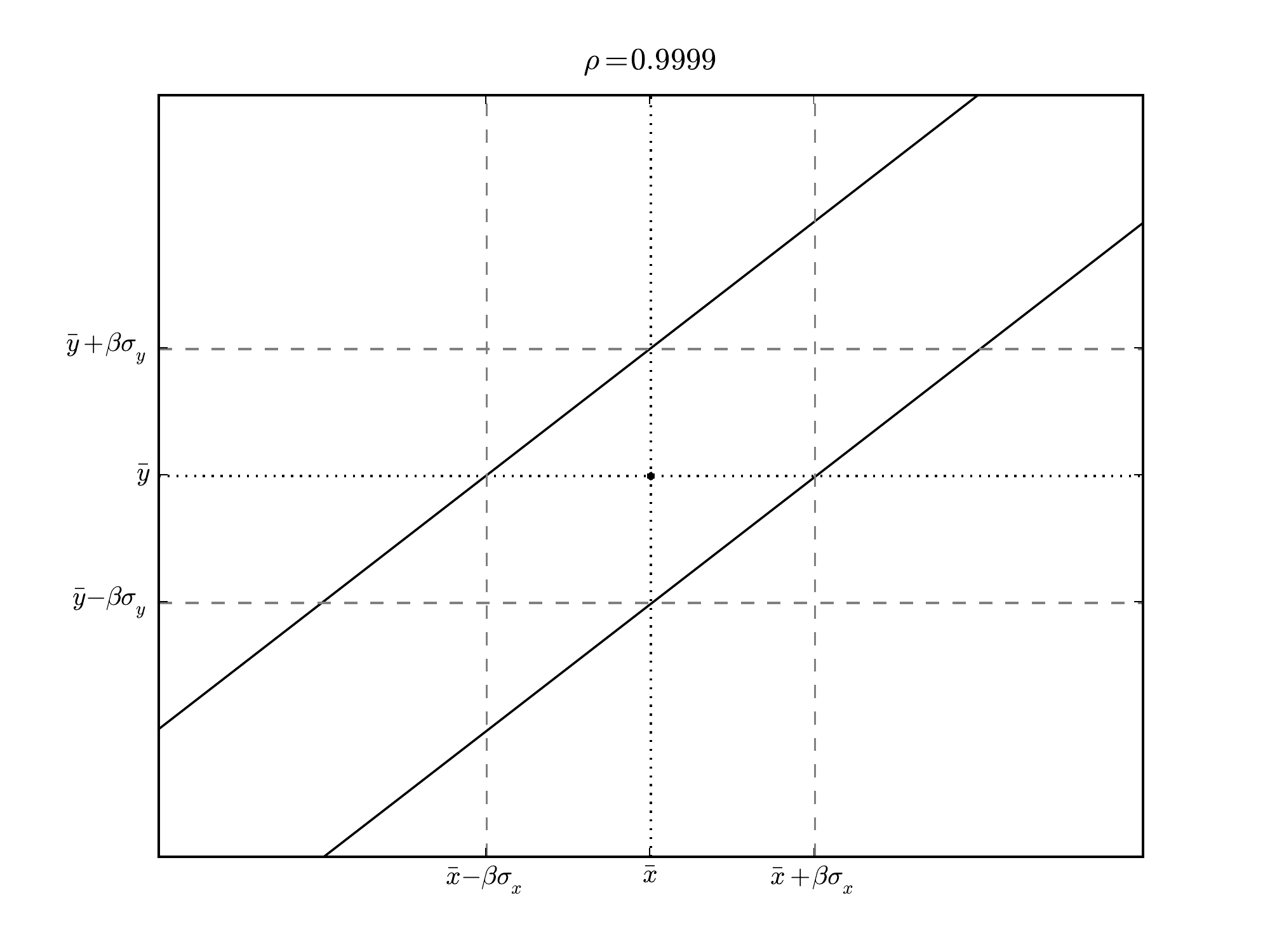}
\caption[]{
\label{ellplot}%
68.3\% (1-$\sigma$) confidence ellipses for parameters $x$ and $y$
with 1-$\sigma$ uncertainties $\sigma_x$ and $\sigma_y$
and correlation coefficient $\rho$.
In the first three panels, we plot as dashed lines
the marginalized 1-$\sigma$ uncertainty for each variable:
$\alpha \sigma_x$ and $\alpha \sigma_y$,
where $\alpha \approx \sqrt{2.3} \approx 1.52$.
In the bottom-right panel, 
we zoom in to show the intersections with the axes:
$\pm \beta \sigma_x$ and 
$\pm \beta \sigma_y$,
where $\beta \approx 2.13 \sqrt{1-\rho}$ (for $\rho \approx 1$).
}
\end{figure*}

The inverse of the Fisher matrix is the covariance matrix:
\\
\begin{equation}  %
  \label{eq:Finv}
  \left [ F \right ]^{-1} = 
  \left [ C \right ] =
  \left[
    \begin{array}{cc}
      \dx^2 & \dxy \vspace{0.07in}\\
      \dxy & \dy^2
    \end{array}
  \right]
\end{equation}
\\
$\sigma_x$ and $\sigma_y$ are the 1-$\sigma$ uncertainties
in your parameters $x$ and $y$,
respectively (marginalizing over the other).
$\dxy = \rho \dx \dy$, where
$\rho$ is known as the correlation coefficient.
$\rho$ varies from 0 (independent) to 1 (completely correlated).
Examples are plotted in Fig.~\ref{ellplot}.

The ellipse parameters are calculated as follows \citep[e.g.,][]{ellipses}:
\\
\begin{equation}  %
  a^2 = \frac{\dx^2 + \dy^2}{2}
  + \sqrt{ \frac{(\dx^2 - \dy^2)^2}{4} + \dxy^2 }
\end{equation}
\begin{equation}  %
  b^2 = \frac{\dx^2 + \dy^2}{2}
  - \sqrt{ \frac{(\dx^2 - \dy^2)^2}{4} + \dxy^2 }
\end{equation}
\begin{equation}  %
  \tan 2 \theta = \frac{2 \dxy}{\dx^2 - \dy^2}
\end{equation}
\\
We then multiply the axis lengths $a$ and $b$
by a coefficient $\alpha$
depending on the confidence level we are interested in.
For 68.3\% CL (1-$\sigma$), 
$\Delta \chi^2 \approx 2.3$,
$\alpha = \sqrt{\Delta \chi^2} \approx 1.52$.
Other values can be found in Table \ref{confell}.
These can be calculated following e.g., \cite{Lampton76}.

The area of the ellipse is given by
\\
\begin{eqnarray}
  \label{eq:area}
  A & = & \pi (\alpha a) (\alpha b)\\
  &=& \pi (\Delta \chi^2) a b\\
  &=& \pi \sigma_x \sigma_y \sqrt{1 - \rho^2}\label{eq:area3}
\end{eqnarray}
\\
The inverse of the area is a good measure of figure of merit.
The Dark Energy Task Force \citep[DETF;][]{DETF06,DETF09}
used ${\rm FOM} = \pi / A$ 
for the ability of experiments (WL, SN, BAO, CL)
to constrain the dark energy equation of state parameters ($w_0, w_a$).

\subsection{Probability $P(x,y)$}

Interested in the probability that specific values are correct for 
parameters $x$ and $y$?
The probability function $P(x,y)$
given best fit values $(x_0, y_0)$
and 1-$\sigma$ uncertainties $(\dx, \dy)$ is calculated as follows:
\\
\begin{equation}  %
  \chi^2
  = \frac{ \left( \Dxdx \right)^2 
  + \left( \Dydy \right)^2
  - 2 \rho \left( \Dxdx \right) \left( \Dydy \right)}{1 - \rho^2}
\end{equation}
\begin{equation}  %
  P(x,y) = \exp\left(-\frac{\chi^2}{2}\right)
\end{equation}
\\
with $\Dx \equiv x - x_0$ and $\Dy \equiv y - y_0$.
Note for $\rho = 0$ (uncorrelated $x$ and $y$),
the $\chi^2$ formula looks familiar.
For correlated $x$ and $y$ ($\rho > 0$), $\chi^2$ is reduced.

\begin{deluxetable}{cccc}
\tablecaption{Confidence Ellipses: 
\label{confell}}
\tablehead{
\colhead{$\sigma$} & 
\colhead{CL} & 
\colhead{$\Delta \chi^2$} & 
\colhead{$\alpha$}
}
\startdata
1 & 68.3\% & ~2.3~ & 1.52 \\
2 & 95.4\% & ~6.17 & 2.48 \\
3 & 99.7\% & 11.8~ & 3.44 \\
\vspace{-0.1in}
\enddata






\end{deluxetable}

\section{Manipulation: Marginalization, Priors, Adding Data Sets, and More}
\label{S:manipulation}

Consider a Fisher matrix provided by the DETF
(Table \ref{tab:fish})
for optimistic Stage IV BAO observations
for the following variables:
($\om, \OL, \Ok$),
where $\om \equiv \Om h^2$ and $\Om + \OL + \Ok = 1$.
The covariance matrix (inverse of the Fisher matrix)
is given in Table \ref{tab:cov}.
For example, the top-left element tells us that
$\Delta \om \approx 0.00566 \approx \sqrt{3.20E-5}$.

\begin{deluxetable}{rrrr}
\tablecaption{Example Fisher Matrix \label{tab:fish}}
\tablehead{
&
\colhead{$\om$} & 
\colhead{$\OL$} & 
\colhead{$\Ok$}
}
\startdata
$\om$ & 2,376,145 & 796,031 & 615,114 \\
$\OL$ &   796,031 & 274,627 & 217,371 \\
$\Ok$ &   615,114 & 217,371 & 178,014 \\
\vspace{-0.1in}
\enddata
\end{deluxetable}

\begin{deluxetable}{rrrr}
\tablecaption{Corresponding Covariance Matrix\label{tab:cov}}
\tablehead{
&
\colhead{$\om$} & 
\colhead{$\OL$} & 
\colhead{$\Ok$}
}
\startdata
$\om$ & ~3.20E-5 & -1.56E-4 & ~8.02E-5 \\
$\OL$ & -1.56E-4 & ~8.71E-4 & -5.25E-4 \\
$\Ok$ & ~8.02E-5 & -5.25E-4 & ~3.69E-4 \\
\vspace{-0.1in}
\enddata
\end{deluxetable}

\subsection{Marginalization}

When quoting these uncertainties on $\om$,
the other variables ($\OL, \Ok$)
have automatically been marginalized over.
That is, their probabilities have been integrated over:
they have been set free to hold any values
while we calculate the range of acceptable $\om$.

To calculate a new Fisher matrix marginalized over any variable,
simply remove that variable's row and column from the covariance matrix,
and take the inverse of that to yield the new Fisher matrix.

\subsection{Fixing Parameters}

Suppose instead want the opposite: perfect knowledge of a parameter.
For example, we want to consider a flat universe with a fixed value of $\Ok = 0$.
To do this, simply remove $\Ok$ from the Fisher matrix (Table \ref{tab:fishflat}).
The new covariance matrix and parameter uncertainties
are calculated from the revised Fisher matrix.

Alternatively, the on-diagonal element corresponding to that parameter
can be set to a very large value.
For example,
if we set the bottom-right element in Table \ref{tab:fish} to $10^{12}$,
that would correspond to a $10^{-6}$ uncertainty in $\om$, or nearly fixed.
Note that higher values in the Fisher matrix
correspond to higher certainty.

\begin{deluxetable}{rrrr}
\tablecaption{Fisher Matrix with Fixed $\Ok = 0$ \label{tab:fishflat}}
\tablehead{
&
\colhead{$\om$} & 
\colhead{$\OL$}
}
\startdata
$\om$ & 2,376,145 & 796,031 \\
$\OL$ &   796,031 & 274,627 \\
\vspace{-0.1in}
\enddata
\end{deluxetable}

\subsection{Priors}

Rather than fixing a parameter to an exact value,
we may want to place a prior such as $\Delta \Ok = 0.01$ (1-$\sigma$).
In this case, simply add $1 / \sigma^2 = 10^4$
to the on-diagonal element corresponding to that variable
(in this case, the bottom left element).

\subsection{Adding Data Sets}

To combine constraints from multiple experiments,
simply add their Fisher matrices: $F = F_1 + F_2$.
Strictly speaking, 
any marginalization should be performed after the addition.
But if the ``nuisance parameters'' are uncorrelated
between the two data sets,
then marginalization may be performed before the addition.

\section{How to Calculate your Own Fisher Matrices}
\label{S:calc}

Given the badness of fit $\chi^2(x,y)$,
your 2-D Fisher matrix can be calculated as follows:
\\
\begin{equation}  %
  \label{eq:Finv}
  \left [ F \right ] = 
  \frac{1}{2}
  \left[
    \begin{array}{cc}
      \pxx & \pxy \vspace{0.07in}\\
      \pxy & \pyy
    \end{array}
  \right]
  \chi^2
\end{equation}
\\
In other words,
$F_{ij} = \displaystyle \frac{1}{2} \frac{\p \chi^2}{\p p_i \p p_j}$.

These derivatives are simple to calculate numerically:
\\
\begin{equation}  %
  \label{eq:chisq}
  \dchisqxx \approx
  \frac{\chi^2(x_0 + \Dx, y_0) 
    - 2 \chi^2(x_0, y_0) 
    +   \chi^2(x_0 - \Dx, y_0)}{(\Dx)^2}
\end{equation}
\\
\begin{equation}  %
  \dchisqx \approx
  \frac{\chi^2(x_0 + \Dx, y_0) 
    -   \chi^2(x_0 - \Dx, y_0)}{2 \Dx} \\
\end{equation}  %
\begin{equation}  %
  \dchisqxy = \frac{\p \dchisqx}{\p y}
\end{equation}  %

\section{Transformation of Variables}
\label{S:trans}

Suppose we are given a Fisher matrix in terms of variables $p = (x,y,z)$
but we are interested in constraints on related variables $p^\prime = (a,b,c)$.
We can obtain a new Fisher matrix as follows:
\\
\begin{equation}  %
  \label{eq:transform}
  F^\prime_{mn} = \sum_{ij} 
  \frac{\p p_i}{\p p^\prime_m}
  \frac{\p p_j}{\p p^\prime_n}
  F_{ij}
\end{equation}
\\
Let's spell this out explicitly.  
Here is the expression for element $(a,b)$ in the new Fisher matrix:
\\
\begin{eqnarray}
  \label{eq:transform2}
  F^\prime_{ab}
    & = & \frac{\p x}{\p a} \frac{\p x}{\p b} F_{xx}
        + \frac{\p x}{\p a} \frac{\p y}{\p b} F_{xy}
        + \frac{\p x}{\p a} \frac{\p z}{\p b} F_{xz} \\
  ~ & + & \frac{\p y}{\p a} \frac{\p x}{\p b} F_{yx}
        + \frac{\p y}{\p a} \frac{\p y}{\p b} F_{yy}
        + \frac{\p y}{\p a} \frac{\p z}{\p b} F_{yz} \\
  ~ & + & \frac{\p z}{\p a} \frac{\p x}{\p b} F_{zx}
        + \frac{\p z}{\p a} \frac{\p y}{\p b} F_{zy}
        + \frac{\p z}{\p a} \frac{\p z}{\p b} F_{zz}
\end{eqnarray}
\\
This can be calculated using matrices:
\\
\begin{equation}  %
  \label{eq:transform3}
  [F^\prime] = [M]^T [F] [M]
\end{equation}
\\
where $M_{ij} = \displaystyle \frac{\p p_i}{\p p^\prime_j}$:
\\
\begin{equation}  %
  \label{eq:Finv}
  \left [ M \right ] = 
  \left[
    \begin{array}{ccc}
      \dxa & \dxb & \dxc \vspace{0.1in}\\
      \dya & \dyb & \dyc \vspace{0.1in}\\
      \dza & \dzb & \dzc
    \end{array}
  \right]
\end{equation}
\\
and $[M]^T$ is the transpose.

All of these partial derivatives should be evaluated numerically,
plugging in best-fit values of the parameters.

\subsection{Transformation Example}

Suppose we are given a Fisher matrix in terms of ($\om, \OL, \Ok$),
but we are interested in ($\Om, \OL, h$).
Here $\om \equiv \Om h^2$ and $\Ok = 1 - \Om - \OL$.
Suppose further that the best-fit cosmology is 
$(\Om, \OL, h) = (0.3, 0.7, 0.7)$.
Our transformation matrix is evaluated as follows:
\\
\begin{eqnarray}  %
  \label{eq:transex}
  \left [ M \right ] 
  & = &
  \left[
    \begin{array}{ccc}
      \domOm & \domOL & \domh \vspace{0.1in}\\
      \dOLOm & \dOLOL & \dOLh \vspace{0.1in}\\
      \dOkOm & \dOkOL & \dOkh
    \end{array}
  \right]\\\\
  ~ & = &
  \left[
    \begin{array}{ccc}
      h^2 &  0 & 2\Om h \vspace{0.1in}\\
       0  &  1 & 0 \vspace{0.1in}\\
      -1  & -1 & 0
    \end{array}
  \right]
  =
  \left[
    \begin{array}{ccc}
      0.49 &  0 & 0.42 \vspace{0.1in}\\
       0  &  1 & 0 \vspace{0.1in}\\
      -1  & -1 & 0
    \end{array}
  \right]
\end{eqnarray}
\\

\section{Pivot Redshift}
\label{S:zp}

Given the dark energy equation of state parameterization
\\
\begin{equation}  %
  \label{eq:w}
  w = w_0 + (1 - a) w_a
\end{equation}
\\
where $1 / a = 1 + z$,
if you have calculated a Fisher Matrix for dark energy parameters $w_0$ and $w_a$,
go ahead and calculate the pivot redshift, too:
\\
\begin{equation}  %
  \label{eq:pivot}
  z_p = \frac{-1}{1 + \frac{\displaystyle \Delta w_a}{\displaystyle \rho \Delta w_0}}
\end{equation}
\\
At this redshift, $w(z)$ is best constrained \citep[e.g., Fig.~16 of][]{HutererTurner01}.
Rather than presenting constraints on ($w_0, w_a$),
constraints on ($w_p, w_a$) can be presented.
That is, we constrain the value of $w$ at $z = z_p$ rather than at $z = 0$
(along with $w$'s rate of change with time $w_a$).

The ($w_p, w_a$) confidence ellipse has no tilt;
there is no correlation between the two, by definition.\footnote{Thus the DETF chooses
a more interesting ellipse to plot: ($w_p, \Omega_{DE}$).}
But the area of the ($w_p, w_a$) ellipse 
is equal to the area of the ($w_0, w_a$) ellipse.
From this and Eq.~\ref{eq:area3} it follows that
\\
\begin{equation}  %
  \label{eq:dwpw0}
  \Delta w_p = \Delta w_0 \sqrt{1 - \rho^2}
\end{equation}
\\
And if $w$ is constant, then $\Delta w_p = \Delta w_0$.

Derivation of the pivot redshift formula 
follows from \citep{DETF06},
calculating the uncertainty of
\\
\begin{equation}  %
  \label{eq:wp}
  w_p = w_0 + (1 - a_p) w_a
\end{equation}
\\
\\
\begin{equation}  %
  \label{eq:dwp}
  ( \Delta w_p )^2 = ( \Delta w_0 )^2 + ( (1 - a_p) \Delta w_a )^2 + 2 (1 - a_p) \Delta w_{0,a}
\end{equation}
\\
where $\Delta w_{0,a} = \rho \Delta w_0 \Delta w_a$,
and then minimizing $\Delta w_p$ for $a_p$.

\section{Discussion}
\label{S:disc}

Fisher matrices encode the Gaussian uncertainties in a number of parameters.
Confidence ellipses can be easily calculated over any pair of parameters.
These provide an optimistic approximation to the true probability distribution.
The true uncertainties may be larger and non-Gaussian.
Note the best fit values themselves are not encoded in the Fisher matrices,
and must be provided separately.

Fisher matrices allow one to easily manipulate 
parameter constraints over many variables.
It is easy to add data sets, add priors, marginalize over parameters,
and transform variables, as shown here.

A more in-depth discussion of Fisher matrices and issues surrounding their use
can be found in \citep{DETF09}.

This is the paper I'd wished I could find
when I began my work with Fisher matrices:
projections for cosmological constraints from gravitational lens time delays
\citep{CoeMoustakas09b}.

\section{Software}
\label{S:software}

\textbf{Fisher.py}\footnote{\href{http://www.its.caltech.edu/\%7Ecoe/Fisher/}{\tt  http://www.its.caltech.edu/\%7Ecoe/Fisher}/} Python -- 
simple manipulation of Fisher matrices and plotting of ellipses

\textbf{DETFast}\footnote{\href{http://www.physics.ucdavis.edu/DETFast/}{\tt http://www.physics.ucdavis.edu/DETFast/}}
\citep{DETF06} 
JAVA -- Compare expectations of cosmological constraints from different experiments with your choice of priors with a few clicks!

\textbf{Fisher4Cast}\footnote{\href{http://www.cosmology.org.za/}{\tt http://www.cosmology.org.za/}} 
\citep{Fisher4Cast} 
Matlab -- most sophisticated

Your ad here.

\section{Contribute}
\label{S:contribute}

This is meant to be a brief guide, 
but if I've failed to reference another useful guide or your software
or if I've neglected some detail (subtle or otherwise) about Fisher matrices,
please e-mail me at coe(at)caltech.edu, and I'll be happy to update this document.
Also please tell me if any section is unclear.

If I have not covered a useful topic,
it is probably outside my knowledge of Fisher matrices.
For example, I have not covered
the analysis of Monte Carlo Markov Chains (MCMC)
as provided, for example, by the WMAP Lambda website.\footnote{
\href{http://lambda.gsfc.nasa.gov/}{\tt http://lambda.gsfc.nasa.gov/}}
If a generous reader could explain to me
(or point me to an appropriate reference on)
how to extract confidence contours and a Fisher matrix from a MCMC,
I would be grateful and include the explanation here,
giving due credit to the contributor.

\acknowledgements

I thank Olivier Dore for referring me to the DETFast software
written by Jason Dick and Lloyd Knox
whom I also thank for answering my questions about their software.
It is a valuable resource.
Once I took off these training wheels and began to produce my own plots,
DETFast is still a valuable resource
for Fisher matrices calculated by the DETF
encoding their estimates of cosmological constraints from various future experiments.

This work was carried out at Jet Propulsion Laboratory,
California Institute of Technology, under a contract with NASA.

\bibliographystyle{astroads}
\bibliography{paperstrunc}

\end{document}